\documentclass{jpsj2}

\title{The nanoscale phase separation in hole-doped manganites}

\author{Roland \textsc{Mathieu}$^{1,2}$\thanks{Now at:  Department of Microelectronics and Applied Physics, Condensed Matter Physics group (KMF), Royal Institute of Technology (KTH), Electrum 229, SE-164 40 Kista, Sweden. Electronic address: rmathieu@kth.se} and Yoshinori \textsc{Tokura}$^{1,3,4,5}$}

\inst{$^{1}$ Spin Superstructure Project (ERATO-SSS), Japan Science and Technology Agency (JST), AIST Central 4, Tsukuba 305-8562, Japan\\
$^{2}$ Department of Advanced Materials Science, University of Tokyo, Kashiwa, Chiba 277-8561, Japan, and CREST, JST, Kawaguchi, Saitama 332-0012, Japan\\
$^{3}$ERATO multiferroics project, JST, c/o Department of Applied Physics
, University of Tokyo, Tokyo 113-8656, Japan\\
$^{4}$Department of Applied Physics, University of Tokyo, Tokyo 113-8656,
 Japan\\
$^{5}$Correlated Electron Research Center (CERC), AIST Central 4, Tsukuba
 305-8562, Japan}

\abst{A macroscopic phase separation, in which ferromagnetic clusters are observed in an insulating matrix, is sometimes observed, and believed to be essential to the colossal magnetoresistive (CMR) properties of manganese oxides. The application of a magnetic field may indeed trigger large magnetoresistance effects due to the percolation between clusters allowing the movement of the charge carriers. However, this macroscopic phase separation is mainly related to extrinsic defects or impurities, which hinder the long-ranged charge-orbital order of the system. We show in the present article that rather than the macroscopic phase separation, an homogeneous short-ranged charge-orbital order accompanied by a spin glass state occurs, as an intrinsic  result of the uniformity of the random potential perturbation induced by the solid solution of the cations on the $A$-sites of the structure of these materials. Hence the phase separation does occur, but in a more subtle and interesting nanoscopic form, here referred as ``homogeneous''. Remarkably, this ``nanoscale phase separation'' alone is able to bring forth the colossal magnetoresistance in the perovskite manganites, and is potentially relevant to a wide variety of other magnetic and/or electrical properties of manganites, as well as many other transition metal oxides, in bulk or thin film form as we exemplify throughout the article.}

\kword{manganites, single-layered manganites, charge-orbital order, magnetic order, electronic phase diagrams, quenched disorder, spin-glasses, phase transitions}

\begin{document}
\maketitle

\section{Introduction}

Oxides with strong electronic correlation exhibit a wide variety of magneto-transport phenomena, due to the srtong coupling among spin, charge, orbital, and lattice degrees of freedoms\cite{orbital,dagotto-science}. For example, the observation of the colossal magnetoresistance\cite{orbital,dagotto-science} (CMR) effect in hole-doped $R_{1-x}A_x$MnO$_3$ manganites ($R$ and $A$ being rare earth and alkaline earth ions, respectively, on the $A$-site of the perovskite $AB$O$_3$ structure. Mn occupies the $B$-site) is not only determined by the one-electron bandwidth $W$, but also by the degree of quenched disorder\cite{orbital,Akahoshi}. Depending on the radii of the $R$ and $A$ cations, the ferromagnetic (FM) metallic phase, or the charge- and orbital-ordered (CO-OO) phase can be stabilized\cite{tomioka-diag}. This CO-OO state, which is associated with the so-called CE-type spin ordering\cite{Jirak}, is schematically illustrated in Fig.~\ref{fig1-CEtype}. This structure is essentially composed of ferromagnetic zig-zag chains antiferromagnetically coupled to one-another. A fragment of such a zig-zag chain is highlighted in the figure. In the half-doped $R_{0.5}$Ba$_{0.5}$MnO$_3$  ($R$ being a rare earth cation), the charge-orbital ordered insulating (favored by small $R$ cations like Y, Gd, or Eu) and ferromagnetic (FM) metallic (larger $R$ cations like Pr or La) states compete with each other, and bi-critically meet near $R$ = Nd\cite{Akahoshi}. In the presence of quenched disorder, namely when the perovskite $A$-sites are solid solution of $R$ and Ba, the phase diagram becomes asymmetric. The FM phase transition is still observed near the critical point, even though the Curie temperature $T_c$ is steeply diminished. The long-range CO-OO state is, on the other hand, completely suppressed and only short-range CO-OO correlation is observed. This phase corresponds in the spin sector to a spin glass (SG) state, which, as we will show in this article, is not related to some macroscopic phase separation\cite{Raveau,Kimura}, but results from the frustration and magnetic disorder microscopically introduced within this ``CE-glass''\cite{Dagotto} state, or nanoscale phase separation.\\
\indent In this article, we sum up the results of our recent studies on high-quality single-crystals of hole-doped manganites, in order to present a comprehensive picture of the nanoscale phase separation occuring in manganites and other transition metal oxides. We also clarify the exact nature of spin-glass phases, as key information can be extracted from the rigorous analysis of their static and dynamical properties. We first examine the nanoscale phase separation in pseudocubic perovskite manganites showing the CMR effect, to compare their properties to those of single-layered manganites with reduced electronic bandwidth. The effect of the hole concentration on the spin- and charge-orbital order is discussed in the single-layered manganites with weak quenched disorder. The homogeneous spin-glass phase which is characterized here in detail also appears in the phase diagram of other transition metal oxides. As an example, the striking resemblance between the phase diagrams of hole-doped manganites and Mo pyrochlores is discussed. The dynamical properties of macroscopically phase-separated manganites ($B$-site doped) are presented for comparison, in order to illustrate the lack of spin-glass phase transition associated with the macroscopic inhomogeneity. Final remarks concerning the potential relevance to novel magneto-transport phenomena in transition metal oxides conclude the article.\\

\section{Experimental}

Throughout the article, we use data from single crystals of hole-doped manganites (see below) as well as literature data on Mo pyrochlores\cite{hanasakitrue,hanasaki}. $A$-site disordered $R_{0.5}$Ba$_{0.5}$MnO$_3$  ($R$ = Eu, Sm, Nd, Pr, and La), $R_{0.5}$Ca$_{1.5}$MnO$_4$, $R_{0.5}$Sr$_{1.5}$MnO$_4$, and  $R_{0.5}$(Ca$_{1-y}$Sr$_y$)$_{1.5}$MnO$_4$ ($R$ = La, La$_{1-y}$Pr$_{y}$ , Pr, Nd, La$_{0.5}$Eu$_{0.5}$  ($\sim$ Nd), Sm, or Eu, while $A$=Ca, Ca$_{1-y}$Sr$_y$), Pr$_{1-x}$Ca$_{1+x}$MnO$_4$ with the hole concentration $x$ varying from $x$=0.3 to 0.7 in steps of 0.05 were grown by the floating zone method\cite{Akahoshi,roland-diag,roland-pcmo}. The high-quality and phase-purity of the crystals were checked by x-ray diffraction. We also show new results on the dynamical properties of a $B$-site doped Nd$_{0.5}$Ca$_{0.5}$Mn$_{0.98}$Cr$_{0.02}$O$_3$ single-crystal also grown by the floating zone method\cite{Kimura}; the perfect Cr stoichiometry and homogeneous distribution was confirmed\cite{Kimura}. The magnetization and ac-susceptibility $\chi$($T,\omega=2\pi f$) data were recorded on a MPMSXL SQUID magnetometer equipped with the ultra low-field option (low frequencies) and a PPMS6000 (higher frequencies), after carefully zeroing or compensating the background magnetic fields of the systems. Additional phase corrections were performed for some frequencies. The transport properties were measured using a standard four-probe method.

\section{Discussion}

\subsection{Half-doped perovskite manganites}

The asymmetric phase diagram of the $A$-site disordered $R_{0.55}$Sr$_{0.45}$MnO$_3$ perovskites\cite{tomioka-bi} is illustrated in the upper panel of Fig.~\ref{fig2-diag-rt}. A FM phase is observed for large cations. The Curie temperature $T_c$ is steeply suppressed as the size of the $R$ ion decreases. For smaller $R$ cations, a spin glass phase with non-equilibrium dynamics emerges.  Interestingly, in a system like Pr$_{0.55}$(Ca$_{1-y}$Sr$_y$)$_{0.45}$MnO$_3$, where the mismatch between the sizes of the different $A$-site cations is much smaller, the suppression of the FM phase is not observed. In that case, as illustrated in the upper panel of Fig.~\ref{fig2-diag-rt}, a CO-OO phase is observed for the small cations\cite{tomioka-bi}. The phase diagrams of the $A$-site ordered and disordered $R_{0.5}$Ba$_{0.5}$MnO$_3$ perovskites are shown in the lower panel of Fig.~\ref{fig2-diag-rt}. In absence of disorder,  a bi-critical phase diagram similar to that of Pr$_{0.55}$(Ca$_{1-y}$Sr$_y$)$_{0.45}$MnO$_3$ is observed. However in the disordered case, akin to the case of $R_{0.55}$Sr$_{0.45}$MnO$_3$, $T_c$ is steeply suppressed as the $R$ ion size decreases. For smaller $R$ cations, a spin glass phase is also observed. Theoretically, such an asymmetric phase diagram can be obtained when considering the multi-critical competition between CO-OO and ferromagnetism\cite{Nagaosa}, in the presence of disorder, which affects the CO-OO more strongly than FM\cite{Nagaosa}. The inset of the lower panel of Figure~\ref{fig2-diag-rt} shows the resistivity of the crystals with $R$ = Eu, Nd, and La (disordered $R$/Ba) measured in zero and applied magnetic field. Eu$_{0.5}$Ba$_{0.5}$MnO$_3$ and Sm$_{0.5}$Ba$_{0.5}$MnO$_3$ (not shown) show insulator-like resistivity in 0 and 7 T; larger magnetic fields may be required to induce metallicity and magnetoresistance.

\begin{figure}[h]
\begin{center}
\includegraphics[width=0.5\textwidth]{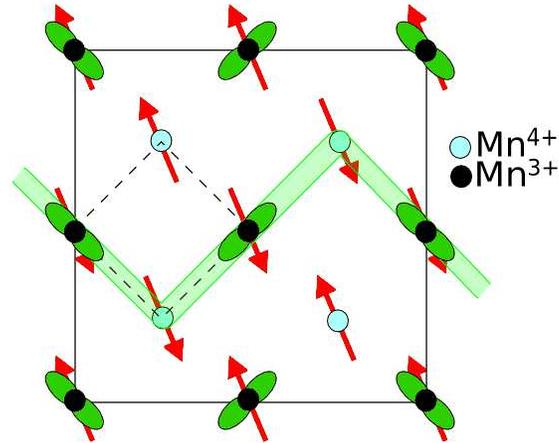}
\end{center}
\caption{Schematic view of the CE-type structure in the basal plane of the tetragonal structure. The orbital order involves staggered $3x^2-r^2/3y^2-r^2$ orbitals of the $e_g$-like electrons of Mn$^{3+}$, represented as green (dark gray) lobes in the figure. The spins, represented with red (dark gray) arrows, order ferromagnetically along zig-zag chains, a fragment of which is highlighted in light green (light gray) in the figure. The dotted square represents the original $I4/mmm$ unit cell, while the filled square represents the magnetic unit cell.}
\label{fig1-CEtype}
\end{figure}

\begin{figure}[h]
\begin{center}
\includegraphics[width=0.66\textwidth]{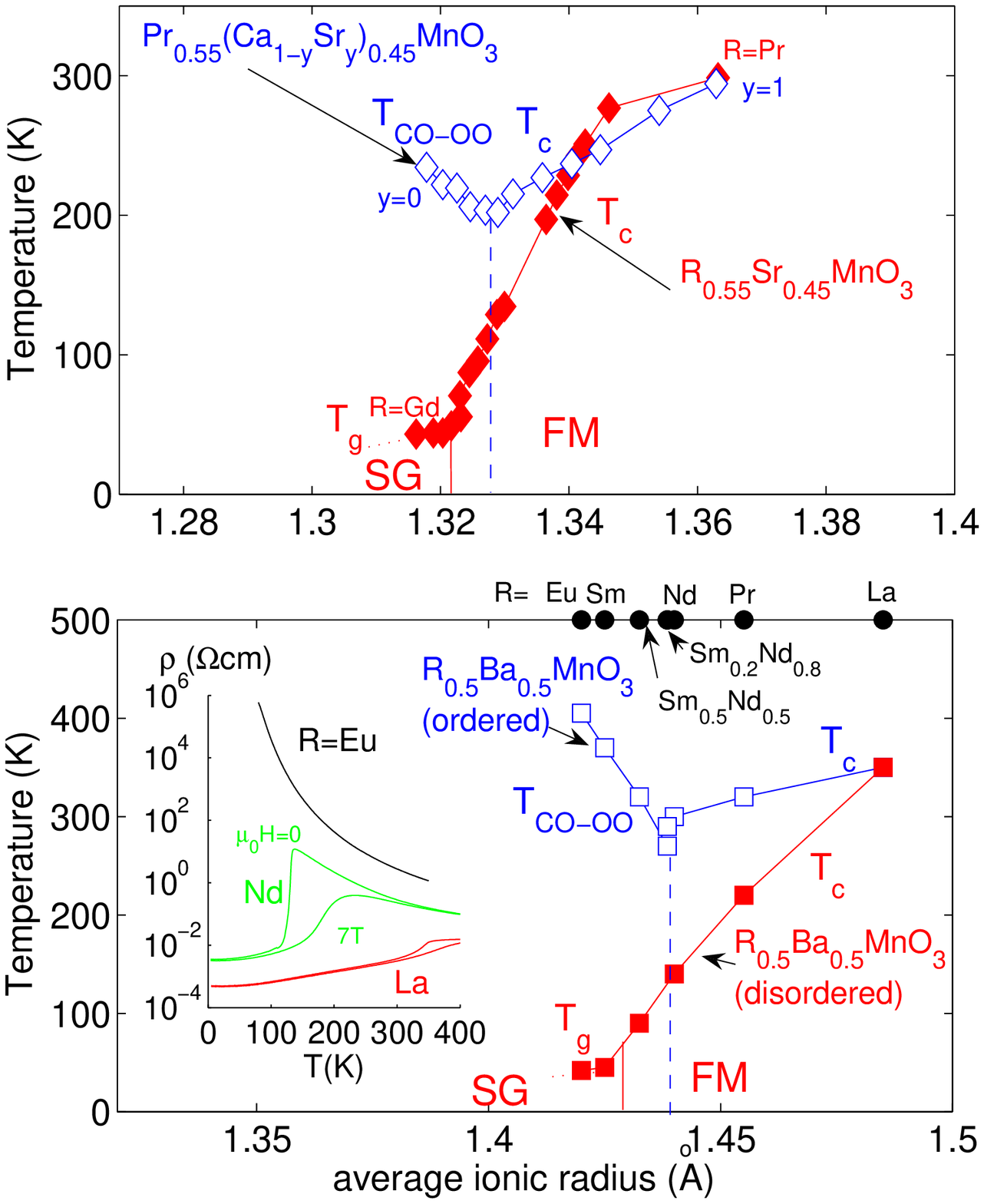}
\end{center}
\caption{Electronic phase diagrams as a function of the average $A$-site ionic radius for Top: Pr$_{0.55}$(Ca$_{1-y}$Sr$_y$)$_{0.45}$MnO$_3$ and $R_{0.55}$Sr$_{0.45}$MnO$_3$. Data adapted from Ref.~\cite{tomioka-bi}; hysteresis around the CO-OO transition and AFM transitions were neglected. Bottom: $A$-site ordered and disordered $R_{0.5}$Ba$_{0.5}$MnO$_3$. Low ionic radii correspond to $R$ = Eu, Sm while the large ones correspond to $R$ = Pr, La. $T_c$ is the Curie temperature,  $T_{\rm co-oo}$ the charge-orbital ordering temperature, and $T_g$ the spin glass (SG) phase transition obtained from dynamical scaling (see main text). The inset shows the temperature dependence of the resistivity of the crystals with  $R$ = Eu, Nd, and La, measured under $\mu_0H$ = 0 and 7 T. See Ref.~\cite{roland-ebmo} for details.}
\label{fig2-diag-rt}
\end{figure}

\noindent However, a FM metallic state can be induced by application of hydrostatic pressure\cite{Takeshita}. $R_{0.5}$Ba$_{0.5}$MnO$_3$ samples with larger $R$ cations occupy a ``critical zone'' near the multi-critical point of the phase diagram. For example, Nd$_{0.5}$Ba$_{0.5}$MnO$_3$ shows a metallic ground state, as well as a large CMR effect in the vicinity of $T_c$.\\
It is known that the long-range CO-OO order can be locally hindered by impurities substituting the Mn-sites, as in Cr doped Pr$_{0.5}$Ca$_{0.5}$MnO$_3$\cite{Raveau} or Nd$_{0.5}$Ca$_{0.5}$MnO$_3$\cite{Kimura}. For Cr concentrations $\leq$ 5 \%, the CO-OO coherence remains on relatively large length scales, yielding separation of FM and CO-OO phases\cite{Kimura}, and associated percolative metal-to-insulator transition upon field application. In the present case, on the contrary, the $R$/Ba solid solution on the perovskite $A$-sites induces a global randomness in the potential, which breaks the CO-OO coherence down to the nanometer scale.The short-range nature of the charge-orbital correlation has been evidenced by the x-ray diffuse scattering observed at all temperatures in Eu$_{0.5}$Ba$_{0.5}$MnO$_3$\cite{roland-ebmo}.

\begin{table}[h]
\caption{Scaling relations for the observation time $\tau$ and  out-of-phase component of the ac-susceptibility $\chi''$=$\chi''$($T,\omega=2\pi f$) in case of critical slowing down and simple thermal activation.  $\epsilon=(T_f(f)-T_g)/T_g$ is the reduced temperature defined in the main text, $T_g$ is the spin-glass phase transition temperature, while $z$, $\nu$, $\beta$, and $\psi$ are critical exponents. $F$ is a functional form; 3$D$ and 2$D$ refer to three-dimensional and two-dimensional respectively.}
\label{newtable}
\small
\vspace*{0.5cm}
\begin{tabular}{c|c|c|c} 
\multicolumn{4}{l}{Critical slowing down (e.g. SG phase transitions):}\\
3$D$ ($T_g$ $\neq$ 0 K) &$\frac{\tau}{\tau_0} = \epsilon^{-z\nu}$&$\chi''T\epsilon^{-\beta}=F(2\pi f \tau_0\epsilon^{-z\nu})$ &See Ref. \cite{Petra}\\
\hline
2$D$ ($T_g$ $=$ 0 K)&$log(\frac{\tau}{\tau_0})\propto \frac{1}{T_f(f)^{1+\psi\nu}}$&$\chi''T^{-\psi\nu}=F(-T^{1+\psi\nu}log(2\pi f \tau_0))$&See Ref.  \cite{roland-2d}\\
\multicolumn{4}{l}{Simple thermal activation (e.g. superparamagnetism):}\\
&$log(\frac{\tau}{\tau_0})\propto \frac{1}{T_f(f)}$&$\chi''=F(-Tlog(2\pi f\tau_0))$&See Ref. \cite{superparam}\\
\end{tabular}
\normalsize
\end{table}

%

Eu$_{0.5}$Ba$_{0.5}$MnO$_3$ exhibits dynamical features typical of spin glasses, such as aging, memory, and rejuvenation\cite{roland-ebmo}. These phenomena can be observed employing specific cooling protocols while recording the ac-susceptibility $\chi$($T,\omega$), and explained using a convenient real space picture known as the droplet model\cite{ghost,fisher}. In the droplet model, the slow dynamics is related to the slow rearrangement of domain walls of the SG phase by thermal activation. After a quench from the paramagnetic phase into the spin-glass phase the system is trapped in a random non-equilibrium spin configuration which slowly equilibrates or ages. In ac measurements, one probes the system on a typical time scale $\sim$ 1/$\omega$ (= 1/2$\pi f$, $f$ is the frequency of the ac-excitation) or a length scale $L$(1/$\omega$). The out-of-phase component of the susceptibility $\chi''$ (as well as the in-phase component $\chi'$ decreases with increasing time $t$ at a constant temperature, reflecting the aging process\cite{ghost}. Hence the number of droplet excitations of relaxation time 1/$\omega$ decays with time as equilibrium domains are growing. Obviously, a greater number of active droplets are probed if a low frequency is employed [large $L$(1/$\omega$)]. The aging is also observed if $\chi''$($T,\omega$) is recorded against temperature performing a halt during the cooling. The equilibration occurs at the halt temperature $T_h$. This equilibration is recovered on re-heating, as the spin configurations established during the aging are frozen-in upon cooling below $T_h$, and only affected on short length-scales during the re-heating\cite{ghost}. In other words, the system keeps memory of the larger domains equilibrated during the halt\cite{roland-ebmo}.

The out-of-phase component of the susceptibility $\chi''$($T,\omega$) of Eu$_{0.5}$Ba$_{0.5}$MnO$_3$ exhibits a fairly large $f$-dependence\cite{roland-ebmo}. Each frequency corresponds to an observation time $t_{obs} = 1/\omega$ characteristic of the measurement. One can define from each susceptibility curve a frequency dependent freezing temperature $T_f$($\omega$), below which the longest relaxation time of the system exceeds $t_{obs}$, and the system is out-of-equilibrium.  $\tau$($T_f$) = $t_{obs}$ can be scaled with the reduced temperature $\epsilon=(T_f(f)-T_g)/T_g$ ($T_g$ is the spin glass phase transition temperature) using a conventional critical slowing-down power-law relation depicted in Table~\ref{newtable} (3$D$ case). A good scaling is obtained for $T_g$ = 42 $\pm$ 1 K, $z\nu$ = 8 $\pm 1$ and $\tau_0$ $\sim$ 10$^{-13 \pm 1}$ s, which indicates that the time necessary to reach equilibrium becomes longer and longer when approaching $T_g$, and the relaxation time diverges at $T_g$ as $\tau/\tau_0$ = $\epsilon^{-z\nu}$. $z$ and $\nu$ are critical exponents, and  $\tau_0$ represents the microscopic flipping time of the fluctuating entities, which in the present case is close to that of the microscopic spin flip time (10$^{-13}$ s). The value of the product $z\nu$ is similar to those of ordinary atomic SG\cite{ghost}. This indicates that $R_{0.5}$Ba$_{0.5}$MnO$_3$ crystals with small $R$ cations undergo a true spin glass phase transition, and that the low temperature SG phase is homogeneously disordered, down to the nanometer scale (see below). This SG state could be expected considering that without quenched disorder, the long range CO-OO state consists of ferromagnetic zigzag chains running along the [110] direction (cubic setting) which are coupled antiferromagnetically (so-called CE-type)\cite{Jirak}. Thus, in this uniformly disordered case, the fragmentation of the zigzag chains down to the nanometer scale (as revealed by the x-ray diffuse scattering) causes the mixture of AFM and FM bonds on these near atomic length scales. These CO-OO ordered nano-clusters may interact by means of the dipolar interaction. The nanoscale phase separation brings forth a perfect matrix for the CMR effect to originate in, as illustrated by the gigantic response to magnetic fields of the narrow bandwidth crystals. As shown in Table~\ref{newtable}, the whole $\chi''$($T,\omega$) curves may be scaled in order to confirm the phase transition. The above concerns three-dimensional (3$D$) SG states. In the case of 2$D$ SG, the dynamical slowing down is expressed with the generalized Arrhenius law expressed in Table~\ref{newtable}. The scaling laws related to thermal activation are also listed for comparison. 

One can also complement the above dynamical scaling analysis by a so-called static scaling analysis\cite{roland-esmo,bb}, which evidence the divergence of the non-linear susceptibility $\chi_{nl}$. The lowest order term of $\chi_{nl}$, $\chi_3(T)$, is expected to diverge at $T_g$ as $\chi_3 \propto \epsilon^{-\gamma}$, where $\epsilon = (T-T_g)/T_g$ is again the reduced temperature and $\gamma$ is a critical exponent.  A scaling relation of the non-linear susceptibility with the magnetic field $H$ in the critical region of the form: $\chi_{nl} = H^{2\beta/(\gamma + \beta)} G[\epsilon/H ^{2/(\gamma + \beta)}]$, where $\beta$ is another critical exponent\cite{susuki,barbara}, is expected ($G$ is a functional form). For suitable values of the $\beta$ exponent, a good scaling may be obtained, confirming the phase transition and yielding the estimation of other critical exponents\cite{roland-esmo}.

\begin{figure}[h]
\begin{center}
\includegraphics[width=0.6\textwidth]{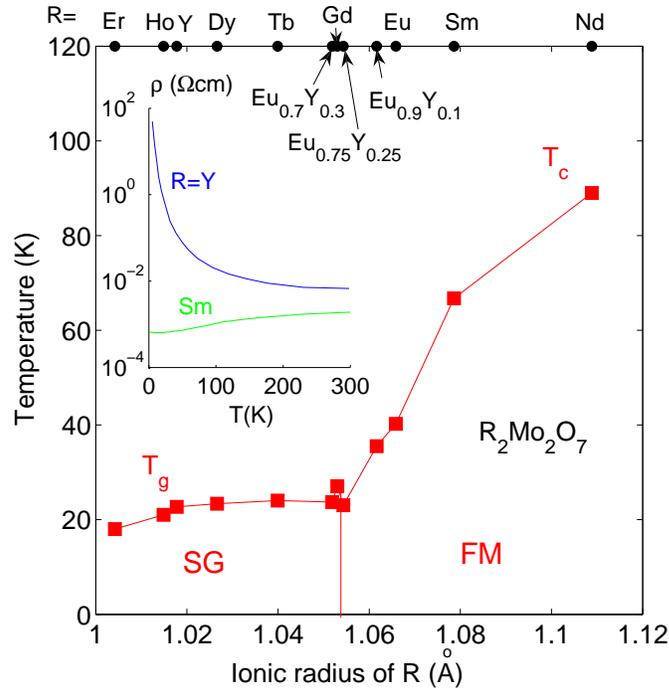}
\end{center}
\caption{Electronic phase diagram for $R_2$Mo$_2$O$_7$ as a function of the ionic radius of $R$. $R$ = Er, Ho, Y, Dy, Tb, Gd, Eu$_{1-y}$Y$_y$, ($y$ = 0.1, 0.3, 0.4), Eu, Sm, and Nd. $T_c$ is the Curie temperature, and $T_g$ the spin glass (SG) phase transition. The inset shows the temperature dependence of the zero-field resistivity of the crystals with  $R$ = Y and Sm. Data adapted from Refs.\cite{hanasakitrue, hanasaki}.}
\label{fig-pyro}
\end{figure}

\subsection{Molybdenum pyrochlores}

Remarkably, the electronic phase diagram of other transition metal oxides resembles that of the hole-doped manganites. For example Mo-pyrochlores ($R_2$Mo$_2$O$_7$) undergo a transition from a ferromagnetic metal to an insulating spin-glass as a function of the size of the $R$ ion, as a result of a similar bi-criticality, in the presence of the geometrical frustration inherent to the pyrochlore lattice\cite{hanasaki,hanasakitrue}. The asymmetric phase diagram of the $R_2$Mo$_2$O$_7$ pyrochlores is depicted in Fig.~\ref{fig-pyro}. The striking similarity with the phase diagrams of the manganites shown in Fig.~\ref{fig2-diag-rt} is evident, with in both cases a SG phase homogeneously disordered down to the nanometer scale. The Mo-pyrochlores with small rare earth ions such as Y$_2$Mo$_2$O$_7$ indeed exhibit dynamical features typical to those of conventional atomic spin-glass, as well as a SG-like phase transition\cite{hanasakitrue,gingras}.  The critical exponent $z\nu$, obtained from the dynamical scaling analyses evidencing the phase transition, amounts to about to $\sim$ 8 for Y$_2$Mo$_2$O$_7$\cite{hanasakitrue}, a value which is similar to the $z\nu$ values of SG systems (See the Table 1 in Ref. \cite{roland-esmo}). This value of $\sim$ 8 is also obtained for other topological spin glass\cite{eric-kagome}. As seen in the inset, the ferromagnetic crystals are metallic, while the spin-glass ones are insulators. 

\subsection{Spin-glasses vs inhomogeneous magnets}

\begin{figure}[h]
\begin{center}
\includegraphics[width=0.6\textwidth]{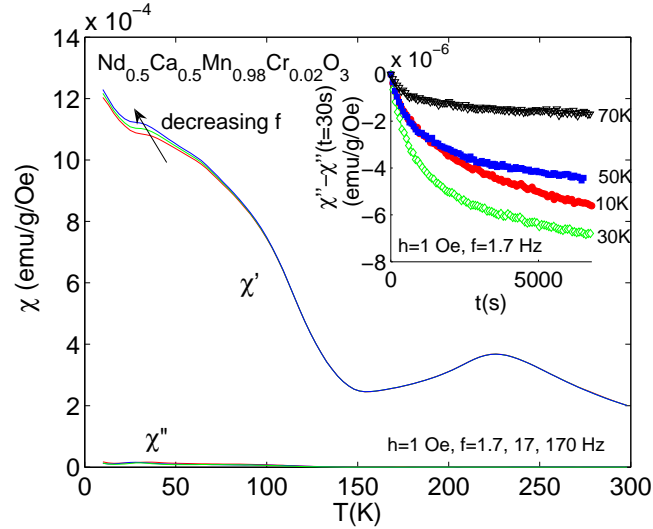}
\end{center}
\caption{Temperature dependence of the in-phase and out-of-phase components of the ac-susceptibility $\chi'$($T,\omega$) and $\chi''$($T,\omega$) for Nd$_{0.5}$Ca$_{0.5}$Mn$_{0.98}$Cr$_{0.02}$O$_3$. The inset shows the time $t$ dependence of the out-of-phase component $\chi''$($T,\omega$) recorded after a quench at selected constant temperatures.}
\label{fig5-ncmocr}
\end{figure}

Any disordered ferromagnet may show glassy features. This shows that this system is magnetically inhomogeneous, but does not however prove that this system undergoes a spin-glass phase transition. For example the micrometer-scale phase separated Nd$_{0.5}$Ca$_{0.5}$Mn$_{0.98}$Cr$_{0.02}$O$_3$, a spin-glass relaxation is observed at low temperatures. This is illustrated in Fig.~\ref{fig5-ncmocr}, which shows how parts of the crystal  exhibit a seemingly long-ranged CO-OO, evidenced by the peak near 230 K, as well as some ferromagnetism, as revealed by the sharp rise of the susceptibility near 150 K. Below this onset of ferromagnetism,  a weak frequency dependence of the susceptibility is observed. As seen in the inset, a spin-glass-like relaxation of the out-of-phase susceptibility can be recorded at low temperatures. However the susceptibility relaxes upwards at higher temperatures, in relation to the ferromagnetic ordering stabilized below 150 K. No scaling analyses similar to the ones depicted above can be performed, evidencing the lack of spin-glass phase transition and thus the microscopic magnetic inhomogeneity of the system.

Frustrated short-ranged magnetic states sometimes evolve at the lowest temperatures into spin-glass states\cite{ycmo}, albeit they usually do not exhibit a spin-glass phase transition. Reentrant ferromagnets, i.e. frustrated ferromagnets which undergo spin-glass phase transitions below their Curie temperatures\cite{kristian}, however essentially behave like ordinary spin-glasses\cite{kristian,roland-epl}. Table~\ref{table1} lists the typical dynamical properties of various simplistic model systems associated with low-temperature glassy phases. For example a typical spin glass system (Type 1 in the table) shows a frequency dependent ac-susceptibility, aging effects in the form of the relaxation of the out-of-phase component of the ac-susceptibility, and a true phase transition as evidenced by (dynamical and/or static) scaling analyzes. The droplet model is sometimes criticized because it predicts that a weak magnetic field will suppress the spin-glass transition, while there are many observations of so-called de Almeida-Thouless lines in various (homogeneous or not) glassy systems describing the evolution of the  spin-glass transition temperature $T_g$ as a function of applied magnetic field ($H-T$ diagrams). It is important to note that in most of these studies, the $T_g$ plotted in the $H$-$T$ diagram has not been obtained from scaling analyses. ``$T_g$'' is often defined as the temperature of the cusp in the zero-field cooled magnetization, or where the zero-field cooled and field cooled magnetization curves start to deviate from each other. This ``$T_g$''  thus represents the temperature onset of irreversibility, and is thus not the real phase transition temperature. Rigorous studies based on the here described scaling analyses show that indeed there is no spin-glass phase transition in a magnetic field, as predicted by the droplet model\cite{atlines}.

Hence it is important to perform such a detailed characterization of the dynamical properties of a system if any meaningful information concerning its glassy properties should be extracted. One should point out that relaxation measurements employing a dc magnetic field can also be performed\cite{ghost,roland-epl}. However it is important to note that in such experiments, the dc magnetic field applied to record the relaxation of the magnetization is only used to record or probe the system (linear response regime), and not induce the aging. An easy demonstration of this is, beside the ``ac memory''  described above, is to perform so-called ``dc-memory'' experiments\cite{roland-dc}. In these experiments, ``memory dips'' similar to the ones obtained in ac experiments, albeit obtained after a halt during the zero-field cooling of the system, are observed on re-heating in a small dc-field. The presence of the dip clearly evidenced the aging of the spin configuration during the initial cooling of the system, which occurred in zero applied magnetic field (as well as its memory). 

\begin{table}[h]
\caption{The dynamical features of different types of glassy systems are exemplified (Types 1 to 6), according to the type of constituting building blocks, the variation in size of these building blocks, the high-temperature magnetic state, and interaction between building blocks. Paramagnets (Type 0a) and superparamagnets (Type 0b) are listed for comparison. The table lists whether there is at low temperature a frequency $f$ dependence of the ac-susceptibility $\chi$, a relaxation of the out-of-phase component of the ac-susceptibility $\chi''$, as well as a phase transition. ``Group of spins'' refers to group of coherent atomic spins. *The low-temperature phase is either a SG or a SSG, depending on the size of the building blocks. Note that reentrant ferromagnets exhibit a $\chi''$ relaxation also in the ferromagnetic phase, albeit the associated equilibration is extremely fragile against temperature changes.; **While experimental results suggest the existence of a phase transition in the ferromagnetic case, there is no definite  answer  yet in this antiferromagnetic case.  Unpublished  results\cite{Per} suggest a zero Kelvin phase transition; see  also  Ref.~\cite{roland-2d}.}
\label{table1}
\scriptsize
{\hspace*{-3cm}
\begin{tabular}{c|c|c|c|c||c|c|c||l}
& Building & Variation in &  High-$T$ & Interaction & $f$-dependent & Relaxation & Phase & \\
Type & block (BB) & size of BBs&  state & between BBs & $\chi$  & of $\chi''$ & transition & \hspace*{0.5cm}  Low-$T$ state \\\hline
	0a & atomic spin & No & PM & No & No & No & No & paramagnet (PM) \\
	0b & group of spins & weak or large & PM & No & Yes & No & No & superparamagnet \\
\hline
	1 & atomic spin & No & PM & Yes & Yes & Yes & Yes & spin glass (SG) \\
	2 & group of spins & weak  & PM & Yes & Yes & Yes & Yes & superspin glass (SSG)\\
	3 & group of spins & large  & PM & Yes & Yes & Yes & No & inhomogeneous\\	
	 &  &   & &  &  &  & & disordered magnet\\
\hline
	4 & atomic spin or & weak & FM & Yes & Yes & Yes & Yes & reentrant \\
	& group of spins & & & & & & & ferromagnet* \\
	5 &\hspace*{0.2cm}" & weak & AFM & Yes & Yes & Yes & No?** & reentrant \\
	 & & & & &  & &  & antiferromagnet \\
	6 &\hspace*{0.2cm}"& large & FM or& Yes & Yes & Yes & No &  inhomogeneous\\
	 & & & AFM & & & & &  disordered magnet
\end{tabular}
\normalsize
}
\end{table}

Also, the fluctuating entities responsible of the present dynamical behavior may comprise more than one single atomic spin. However the total number of spins must be small as the $\tau_0$ obtained from the dynamical scaling is atomic-like ($\sim$ 10$^{-13}$ s). In superspin glasses, weak rejuvenation effects are observed as well\cite{Petra,Petrareview}. However in those interacting magnetic nanoparticle systems, the fluctuating entities are very large, $\sim$ 10$^3$ coherent atomic spins, with a flipping time $\tau_0$ $\sim$ 10$^{-4}$-10$^{-8}$ s. Yet, only the homogeneous (monodispersive in this case) interacting systems undergo a spin-glass phase transition akin to archetypal spin-glasses and Eu$_{0.5}$Ba$_{0.5}$MnO$_3$\cite{Petra,Petrareview}. The inhomogeneous systems no dot show a phase transition, albeit glassy behavior. The non-interacting systems only show superparamagnetism (see Table \ref{table1}). 

The widely employed - but questionable - ``cluster glasses'' term may describe the Type 3 in Table \ref{table1}, or inhomogeneous nanoparticles systems\cite{Petrareview}, as well as the manganites showing the macroscopic phase separation. It is obvious that since the phase diagram of the manganites is bi-critical (CO-OO vs FM) or even multi-critical (CO-OO vs FM vs AFM\cite{Akahoshi}), crystals, and of course polycrystalline samples, may easily show such an inhomogeneous magnetic and/or electronic behavior, even if that is not the true behavior of the clean and stochiometric system.\\

\subsection{Half-doped single-layered manganites}

\begin{figure}[h]
\begin{center}
\includegraphics[width=0.6\textwidth]{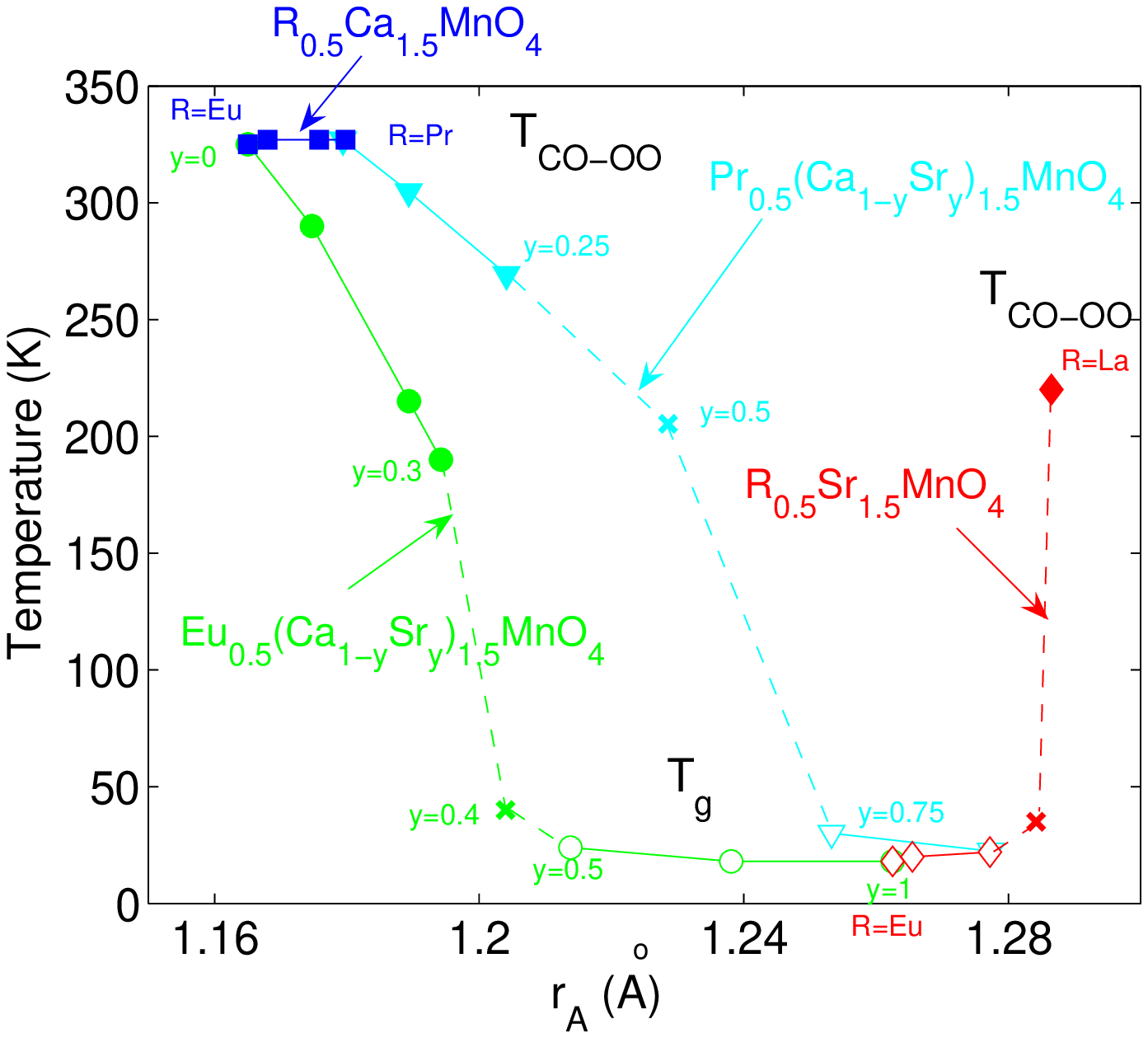}
\end{center}
\caption{Electronic phase diagram showing the evolution of the charge- and orbitally ordered transition temperature $T_{\rm CO-OO}$ (filled symbols) and the spin-glass phase transition temperature $T_g$ (open symbols) as a function of the average $A$-site ionic radius $r_A$ for different single-layered systems ($R_{0.5}$Ca$_{1.5}$MnO$_4$: square markers; $R_{0.5}$Sr$_{1.5}$MnO$_4$: diamonds; Eu$_{0.5}$(Ca$_{1-y}$Sr$_y$)$_{1.5}$MnO$_4$: circles, and  Pr$_{0.5}$(Ca$_{1-y}$Sr$_y$)$_{1.5}$MnO$_4$: triangles). The crosses marks $T_f$ below which the crystals show some glassiness (i.e. no $T_g$). Due to the orbital master - spin slave relationship existing in the half-doped manganites\cite{roland-diag,roland-ebmo}, we here use $T_g$ to reflect the short-ranged nature of the CO-OO, and join the data points for $T_{\rm CO-OO}$, $T_f$ and $T_g$.}
\label{fig4-ra}
\end{figure}

The quenched disorder associated with the solid solution of the $A$-site cations\cite{Akahoshi} can be quantified using the ionic radius variance $\sigma^2=\sum_i x_i r_i^2 - r_A^2$\cite{attfield}. $x_i$ and $r_i$ are the fractional occupancies ($\sum_i x_i$=1) and electronic radii of the different $i$ cations on the $A$-site respectively, and $r_A=\sum_i x_ir_i$ represents the average $A$-site ionic radius\cite{roland-diag}. La$_{0.5}$Sr$_{1.5}$MnO$_4$ is a well known layered manganite with concomitant charge and orbital orders\cite{lsmo-orb} near 220K. The spin sector orders antiferromagnetically at $T_N$=110K\cite{lsmo-orb}. Akin to the perovskite case, crystals with smaller bandwidth (i.e. with smaller $r_A$) such as Pr$_{0.5}$Ca$_{1.5}$MnO$_4$ show CO-OO transitions above room temperature\cite{Raveau}, as shown in Fig.~\ref{fig4-ra}. While the $R_{0.5}$Ca$_{1.5}$MnO$_4$ crystals exhibit a long-ranged CO-OO with the same $T_{\rm CO-OO}$,  $T_{\rm CO-OO}$ decreases rapidly as Sr is introduced in Eu$_{0.5}$(Ca$_{1-y}$Sr$_y$)$_{1.5}$MnO$_4$ and Pr$_{0.5}$(Ca$_{1-y}$Sr$_y$)$_{1.5}$MnO$_4$, until the long-ranged order is not observed. 

\begin{table}[h]
\caption{Hole-doping $x$, average ionic radius ($r_A$), ionic variance ($\sigma^2$), charge-orbital ordering ($T_{\rm CO-OO}$) and spin-glass ($T_g$) temperatures, estimates of the CO-OO coherence length $\xi_{\rm CO-OO}$ as well as flipping times $\tau_0$ obtained from dynamical scaling analyzes for selected crystals. *Coordination is 12 in this case (it is 9 for the remaining layered crystals); **$\xi_{\rm CO-OO}$ is estimated from the inverse of the HWHM of the diffraction data. However for these crystals with long-ranged CO-OO order, the HWHM is determined by the resolution limit of the microscope, rather than the actual CO-OO correlation length.}
\label{table2}
\small
\begin{tabular}{c|c|c|c|c|c|c|c}
System& $x$ & $r_A$ ({\AA})& $\sigma^2$ ({\AA}$^2$)& $T_{\rm CO-OO}$ (K) & $T_g$ (K) & $\xi_{\rm CO-OO}$ (nm)& $\tau_0$ (s)\\
\hline
Eu$_{0.5}$Ba$_{0.5}$MnO$_3$ & 0.5 & 1.420*& 0.0361*& -- & 42 &  $\sim$ 2 (XRD, 35K) & $\sim$ 10$^{-13}$\\
\hline
Eu$_{0.5}$Sr$_{1.5}$MnO$_4$ & 0.5 & 1.262 & 0.0067 & -- & 18 & $\sim$ 2 (ED, 20K) & $\sim$ 10$^{-13}$\\
Pr$_{0.5}$Sr$_{1.5}$MnO$_4$ & 0.5 & 1.277 & 0.0032 & -- & 22 & $\sim $ 4 (ED, 20K) & $\sim$ 10$^{-10}$\\
La$_{0.5}$Sr$_{1.5}$MnO$_4$ & 0.5 & 1.286 & 0.00166 & 220 & -- & long-ranged** (ED, 20K) & --\\
Pr$_{0.5}$Ca$_{1.5}$MnO$_4$ & 0.5 & 1.179 & 1.87$\times$10$^{-7}$ & 325 & -- & long-ranged** (ED, 20K) & --\\
\hline
Pr$_{0.6}$Ca$_{1.4}$MnO$_4$ & 0.4 & 1.179 & 2.10$\times$10$^{-7}$ & 270 & 24 & long-ranged** (ED, 80K) & $\sim$ 10$^{-13}$
\end{tabular}
\normalsize
\end{table}

\begin{figure}[h]
\begin{center}
\includegraphics[width=0.64\textwidth]{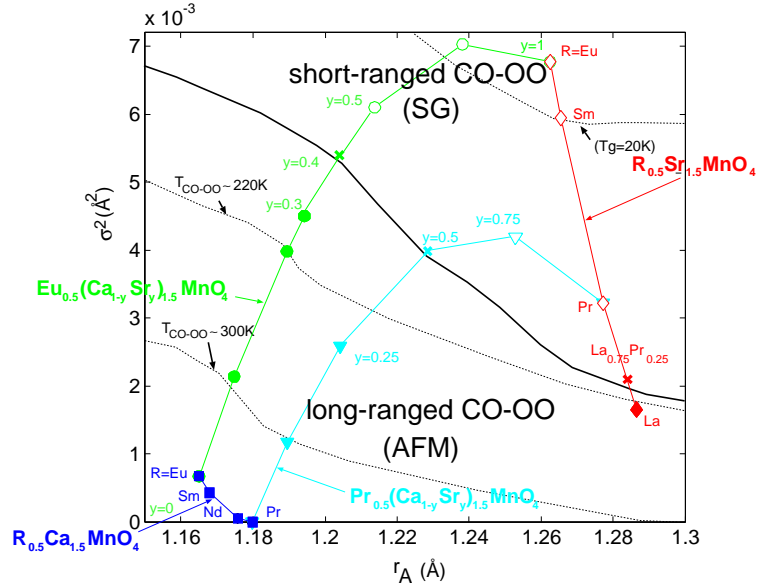}
\end{center}
\caption{Electronic phase diagram of $R_{0.5}$(Ca$_{1-y}$Sr$_y$)$_{1.5}$MnO$_4$ in the plane of the average ionic radius $r_A$ and the variance $\sigma^2$. The data is plotted using the same symbols as in the previous figure for easy comparison; see caption of Fig.\ref{fig4-ra}.}
\label{fig5-bd}
\end{figure}

If $r_A$ increases further, the long-range CO-OO reappears in  La$_{0.5}$Sr$_{1.5}$MnO$_4$ (see Fig.~\ref{fig4-ra}). Hence the short-ranged or long-ranged nature of the CO-OO is not determined only by theaverage ionic radius or the bandwitdh. The degree of quenched disorder is actually much smaller in the $R_{0.5}$Ca$_{1.5}$MnO$_4$ crystals ($\sigma^2$ $\sim$ 2$\times$10$^{-7}$ {\AA}$^2$ for Pr$_{1-x}$Ca$_{1+x}$MnO$_4$), than in La$_{0.5}$Sr$_{1.5}$MnO$_4$ ($\sigma^2$ $\sim$ 10$^{-3}$ {\AA}$^2$). It thus makes sense to draw a global phase diagram in the planes of $r_A$ and $\sigma^2$ to take into account the effects of the variation of both bandwidth and quenched disorder. Such a ``bandwidth-disorder'' phase diagram is drawn in Fig.~\ref{fig5-bd}, using the ac-susceptibility, resistivity, and electron diffraction data, which was found to complement each other\cite{roland-diag}. Indeed a crystal with long-range CO-OO order exhibits (i) a sharp peak in the magnetization or ac-susceptibility at $T_{\rm CO-OO}$ arising from the quenching of the FM spin fluctuation, as well as a tiny inflection marking the associated $T_N$ (ii) clear (and hysteretic in temperature) inflections in the electrical resistivity curves near $T_{\rm CO-OO}$, (iii) sharp superlattice spots below $T_{\rm CO-OO}$ in the electron diffraction (ED) patterns, reflecting the orbital order. In contrast, a crystal with short-range CO-OO order displays: (i) no (or broad if close to the phase boundary) anomaly in the magnetization curves except for a frequency dependent cusp at low temperatures, (ii) no anomaly in the resistivity curves, and (iii) diffuse superlattices spots in the electron or x-ray diffraction, reflecting the short-ranged nature of the orbital order. The remarkable co-variation of these physical properties of the systems is illustrated in table~\ref{table2}. Eu$_{0.5}$Ba$_{0.5}$MnO$_3$ shows short-ranged CO-OO and spin-glass state\cite{roland-ebmo}. A similar behavior is observed for the layered Eu$_{0.5}$Sr$_{1.5}$MnO$_4$. In both cases, the CO-OO correlation length, estimated from the inverse of the CO-OO superlattice spots in the electron- or x-ray diffraction patterns, is nanometer-sized. The flipping time $\tau_o$ estimated from dynamical scalings of the susceptibility data is close to that of a single spin. 

This may indicate that the orbital sector, as a master, controls the spin sector, as a slave,  so that when the CO-OO is homogeneously disordered, the spin sector is also homogeneously disordered. This is confirmed by looking again at Table~\ref{table2}. From Eu$_{0.5}$Sr$_{1.5}$MnO$_4$ and La$_{0.5}$Sr$_{1.5}$MnO$_4$, the quenched disorder decreases, while the bandwidth is nearly constant\cite{uchida}. The  CO-OO correlation length slightly increases from Eu$_{0.5}$Sr$_{1.5}$MnO$_4$ to Pr$_{0.5}$Sr$_{1.5}$MnO$_4$. At the same time, the dynamical scaling of the susceptibility data indicates a larger flipping time $\tau_0$, revealing that the low-temperature fluctuating entities are bigger than in the case of Eu$_{0.5}$Sr$_{1.5}$MnO$_4$. There is thus a clear covariation between the CO-OO correlation length and the size of the "superspins" involved in the SG state. Namely, the increase in CO-OO correlation yields the increase of the superspins building the SG state. These superspins, or groups of coherent spins, may thus be viewed as broken pieces of the CO-OO FM zig-zag chains of the CE-type structure (as represented in Fig.~\ref{fig1-CEtype}). When the CO-OO becomes long-ranged, the spin sector becomes long-ranged, and only an AFM phase transition is observed below $T_{\rm CO-OO}$\cite{uchida}.

The bandwidth-disorder phase diagram of the layered crystal is hence presented in Fig.~\ref{fig5-bd}. This diagram is reminiscent of the diagram obtained for the 3$D$ perovskite case\cite{tomioka-diag} in the small bandwidth area (for larger $W$, FM is observed in the perovskite case). In both cases, the long-range CO-OO order is replaced by a short-range ``CE-glass'' state  (SG state) in the presence of large quenched disorder \cite{roland-ebmo,roland-esmo}. However, the first-order like transition between the CO-OO and CE-glass phases observed in the perovskite case\cite{Akahoshi} does not occur in the layered systems. As indicated by the ED results, the CO-OO correlation length continuously decreases as the quenched disorder increases\cite{uchida}. Such a bandwidth-disorder may actually be ``universal'' in the sense that it may also describe other materials such as nickelates or superconducting cuprates.\\

\begin{figure}[h]
\begin{center}
\includegraphics[width=0.52\textwidth]{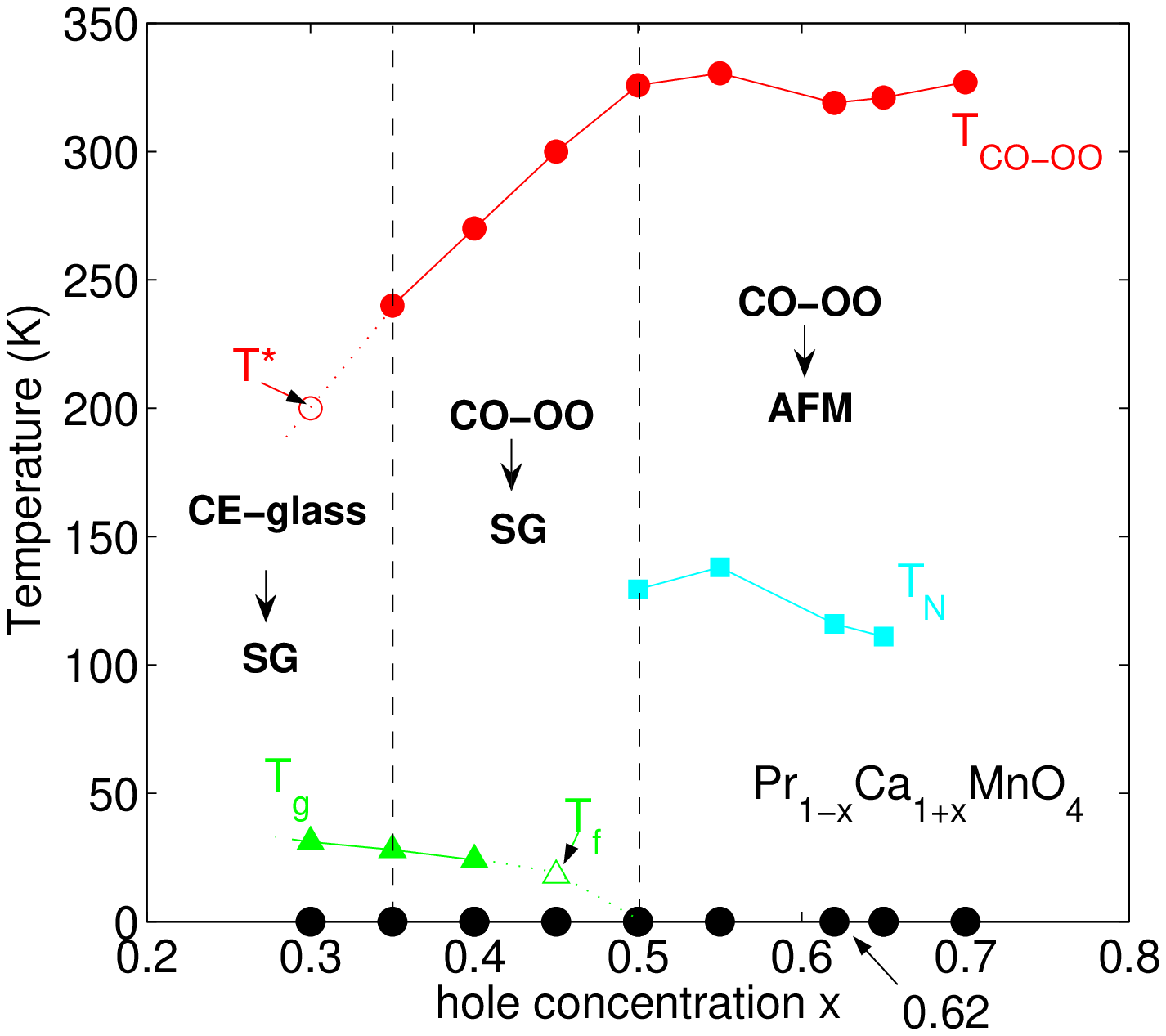}\\
\includegraphics[width=0.52\textwidth]{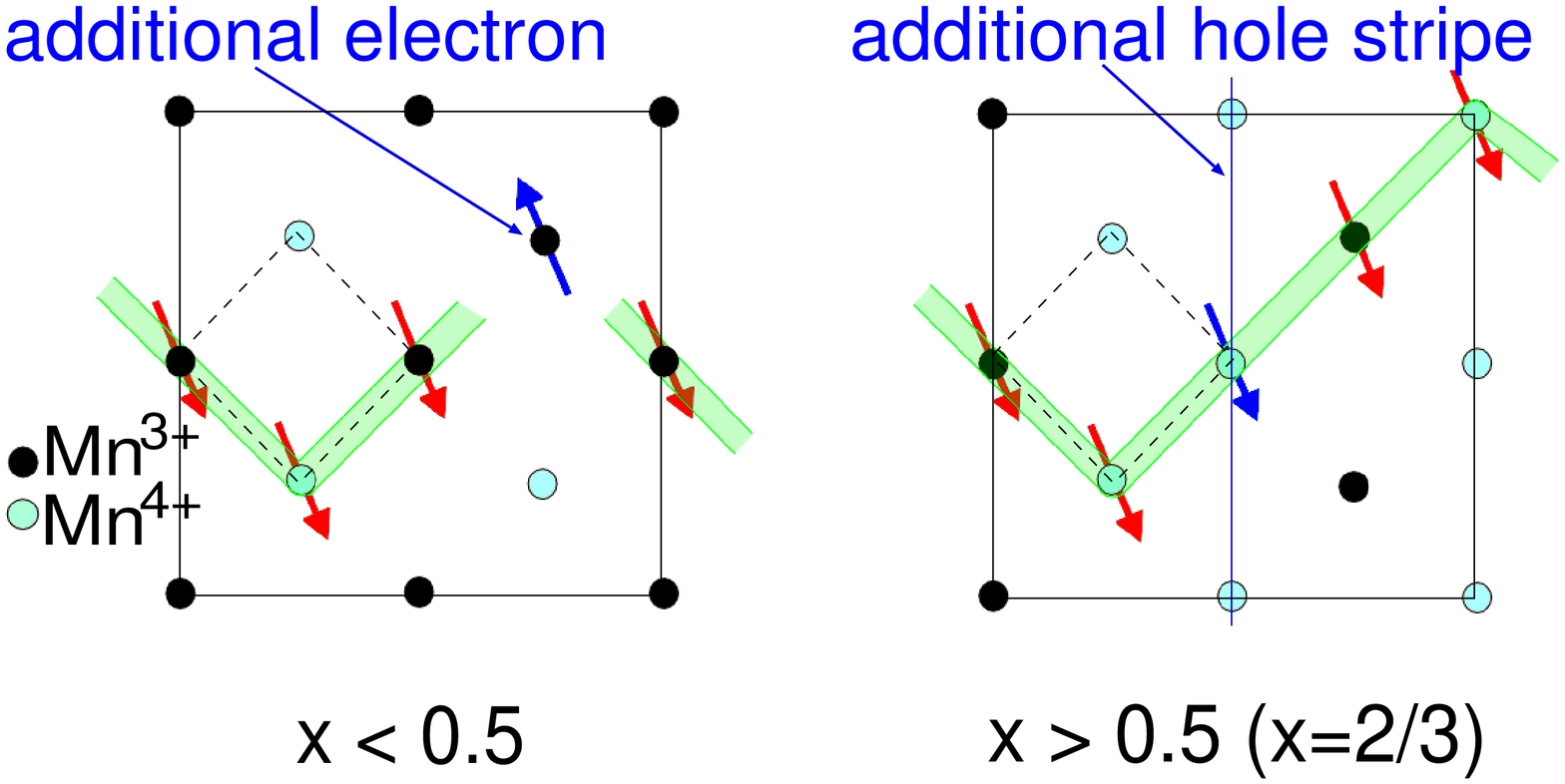}
\end{center}
\caption{Upper panel: Electronic phase diagram for all crystals. The open symbols mark the onsets of the short-ranged orbital ($T^*$, estimated from the ED; see Ref.~\cite{yu}) and magnetic ($T_{\rm S}^*(ab)$) correlation, and of the non-equilibrium dynamics ($T_f$). Data  adapted from R. Mathieu et al., Ref.~\cite{roland-pcmo}. Lower panels: schematic representation of the charge- and orbital ordered states of under-doped ($x<$0.5), and over-doped  ($x>$0.5, namely 2/3) crystals, showing the magnetic arrangement within the zig-zag chains of the CE-type magnetic structure. See Fig.~\ref{fig1-CEtype} for details on the original $x$=0.5 structure.}
\label{fig6-pcmo}
\end{figure}

\subsection{Under- and overdoped single-layered manganites}

The orbital-master spin-slave relationship may appear not to be always true, as it depends on the hole-concentration\cite{fe}. For example the quenched disorder inherent to the solid solution of Pr$^{3+}$/Ca$^{2+}$ ions on the $A$-sites is minimum, as both ions are very similar in size. The degree of disorder does not change significantly with $x$, with $\sigma^2$ only varying from 2.27$\times$10$^{-7}${\AA}$^2$ to 1.27$\times$10$^{-7}${\AA}$^2$ as $x$ varies from 0.3 to 0.7. The half- and over-doped Pr$_{1-x}$Ca$_{1+x}$MnO$_4$ crystals ($x$ $\geq$ 0.5) exhibit sharp CO-OO transitions near $\approx$ 310-330K depending on $x$\cite{roland-pcmo,yu}. In the under-doped crystals, the long-ranged CO-OO remains, albeit with a lower $T_{\rm CO-OO}$ (Fig.~\ref{fig6-pcmo}), until it vanishes below $x$ $\sim$ 0.35. In the underdoped crystals, the ac-susceptibility globally increases as $x$ decreases, and a SG-like frequency dependent cusp appears at low temperatures\cite{roland-pcmo}. Below this peak, a spin-glass-like relaxation is observed, and the dynamical scaling reveals a SG phase transition similar to those of the half-doped crystals\cite{roland-pcmo} (see also Table~\ref{table2}). Hence interestingly, while the orbital sector is long-ranged, the spin sector remains short-ranged, as some frustration is introduced in the spin structure with these additional localized electrons locally affecting the magnetic interaction within the CE zig-zag chains. In the 0.35 $\leq$ $x$ $<$ 0.5 doping range, a frustrated magnetic structure/short-range magnetic correlation is stabilized below $T_{\rm CO-OO}$. The CO-OO correlation length is not finite as in the case of macroscopically phase separated materials. Usually, the above mentioned macroscopic phase separation is observed when defects or impurities  hinder the long-range CO-OO\cite{Raveau,Kimura}. In the above mentioned Cr-doped Nd$_{0.5}$Ca$_{0.5}$MnO$_3$\cite{Kimura}, Cr ions randomly replace Mn sites. Cr$^{3+}$ has the same electronic configuration as Mn$^{4+}$, with no $e_g$ electron.  Cr$^{3+}$ replacing Mn$^{3+}$ strongly affects the CO-OO as the substitution yield an immobile $e_g$ orbital deficiency\cite{Kimura}. However in the present under-doping case, as Mn$^{3+}$ replaces some Mn$^{4+}$, the orbital order is preserved, as only a ``passive'' electron/orbital is added. \\

\section{Conclusions}

To conclude, we have studied $A$-site disordered $R_{0.5}$Ba$_{0.5}$MnO$_3$ pseudocubic perovskite manganites near the spin glass insulator/ferromagnetic metal phase boundary. Analyzes of X-ray diffuse scattering and ac-susceptibility measurements reveal that the crystals with small bandwidth behave like canonical atomic spin glasses, and suggest a short-range orbital order and glassy spin state homogeneous down to the nanometer scale. The random potential spawned by the atomic-scale disorder of the $A$-site cations yields the fragmentation of the zigzag chains on similar length-scales, causing the mixture of AFM and FM bonds on the nanoscale.

The so-called macroscopic phenomenon is not intrinsic, and may be related to the recently introduced concept of ``nanoscale phase separation'' in diluted magnetic semiconductors which may explain the occurrence of ferromagnetism in some doped semiconductors\cite{dietl}. In the manganite case, the macroscopic phase separation occurs only in crystals with local defects or impurities (such as $B$-site dopants), or equivalently, that no phase separation on a micron-scale occurs in high-quality single crystals, when $R$ and Ba are ordered or otherwise perfectly disordered, by no or uniform random potential. 

The nanoscale phase separation is also observed in $A$-site disordered $R_{0.5}A_{1.5}$MnO$_4$ single-layered manganites. However in this case, the gradual decrease of the CO-OO correlation length as a function of bandwidth or disorder occurs, instead of the first-order-like collapse observed in the perovskite case.

Recent theoretical studies also suggest the nanoscale phase separation\cite{Dagotto-new}, and promise to uncover, together with the experimental results, the true building blocks bringing forth the colossal magnetoresistance and associated magnetotransport phenomena\cite{orbital,ziese,rolandbis}. The manganites indeed show many other interesting properties beyond the CMR. For example they exhibit multiferroic properties, from the hexagonal systems\cite{tbo}, to possibly the overdoped single-layered ones, to the double-layered materials\cite{tokunaga}. Hence the understanding, and possible control, of the nanoscale phase separation is essential. This phenomenon is also relevant to device physics and thin films, as heterostructures of different manganites or other oxides\cite{yamada} may show complicated magnetic or electrical (or multiferroic) behavior related to the kind of solid-solution of the different cations at the interfaces. Large resistance switching (sometimes referred to as the colossal electroresistance, CER) at the electrode/crystal\cite{tokunaga2} or electrode/thin film\cite{fujii} interface have also been reported. 

Finally, one should note that the nature of the spin-glass state associated with the nanoscale phase separation itself has a fundamental interest. For example in the above mentioned layered Eu$_{0.5}$Sr$_{1.5}$MnO$_4$ crystal, the anisotropy of the SG state\cite{roland-esmo} gives some insight on the low-temperature orbital state. One may for example speculate that the short-ranged CO-OO state of Eu$_{0.5}$Sr$_{1.5}$MnO$_4$ mainly includes $3x^2-r^2/3y^2-r^2$ orbitals, favoring in-plane magnetic moments. In a related system, namely the electron-doped layered La$_{1.1}$Sr$_{0.9}$MnO$_4$ with an $e_g$ electron occupying the $c$-oriented $d_{3z^2-r^2}$ orbital, the additional electrons introduce a local FM interaction inducing a two-dimensional spin-glass correlation which suggests that the initial antiferromagnetic state of LaSrMnO$_4$ (and that of other layered (La,Sr)MnO$_4$ compositions) is quasi two-dimensional\cite{roland-2d}. We have also here evidenced that interestingly, by growing a bulk single-crystal of a manganite, one can obtain a system behaving dynamically like a monodispersive-like, frozen ferrofluid of magnetic nanoparticles, whose average size can be tuned by the amount of quenched disorder.\\

\section*{Acknowledgment}

The data, analyzes and conclusions presented in this article were obtained within the ERATO Spin Superstructure Project of the Japan Science and Technology Agency (JST), in collaboration with J. P. He, X. Z. Yu, Y. Kaneko, M. Uchida, Y. S. Lee, T. Arima, A. Asamitsu, Y. Matsui, as well as D. Akahoshi, Y. Tomioka, and R. Kumai from the Correlated Electron Research Center (CERC, AIST Tsukuba, Japan), and T. Kimura and N. Hanasaki at the time from the Department of Applied Physics, University of Tokyo, Japan. The valuable discussions with the above researchers, as well as with P. E. J\"onsson, P. Nordblad, and E. Dagotto are gratefully acknowledged.

\end{document}